# Deep learning Architecture for Short-term Passenger Flow Forecasting in Urban Rail Transit

Jinlei Zhang, Feng Chen*, Zhiyong Cui, Yinan Guo, and Yadi Zhu

*Abstract*—Short-term passenger flow forecasting is an essential component in urban rail transit operation. Emerging deep learning models provide good insight into improving prediction precision. Therefore, we propose a deep learning architecture combining the residual network (ResNet), graph convolutional network (GCN), and long short-term memory (LSTM) (called "ResLSTM") to forecast short-term passenger flow in urban rail transit on a network scale. First, improved methodologies of the ResNet, GCN, and attention LSTM models are presented. Then, the model architecture is proposed, wherein ResNet is used to capture deep abstract spatial correlations between subway stations, GCN is applied to extract network topology information, and attention LSTM is used to extract temporal correlations. The model architecture includes four branches for inflow, outflow, graph-network topology, as well as weather conditions and air quality. To the best of our knowledge, this is the first time that air-quality indicators have been taken into account, and their influences on prediction precision quantified. Finally, ResLSTM is applied to the Beijing subway using three time granularities (10, 15, and 30 min) to conduct short-term passenger flow forecasting. A comparison of the prediction performance of ResLSTM with those of many state-of-the-art models illustrates the advantages and robustness of ResLSTM. Moreover, a comparison of the prediction precisions obtained for time granularities of 10, 15, and 30 min indicates that prediction precision increases with increasing time granularity. This study can provide subway operators with insight into short-term passenger flow forecasting by leveraging deep learning models.

## I. INTRODUCTION

Short-term traffic demand forecasting has attracted significant research interest owing to its critical real-world applications. In urban rail transit (URT), for example, short-term passenger flow forecasting (STPFF) can provide real-time traffic information to help passengers make rational scheduling decisions and help transit operators control ridership inflow to avoid congestion, or adjust train timetables to accommodate more passengers in peak hours. Many researchers have devoted considerable effort to studying STPFF in URT. Consequently, STPFF has developed significantly in recent decades.

In the early stage, STPFF development is represented by conventional mathematical statistics-based methods such as the historical average, ordinary least squares, logistic regression, autoregressive integrated moving average (ARIMA), Kalman filter, and *k*-nearest neighbor models. Many researchers have also summarized these models [1]. Because URT did not initially develop so fast that researchers neglected to consider the application of STPFF to URT. However, most of these models are no longer used to analyze road traffic because they cannot meet "real-time" requirements and cannot achieve higher precision than current state-of-the-art models.

With the development of machine learning, some machine-learning-based models and hybrid prediction models are introduced for STPFF, such as backpropagation neural networks (BPNNs) [2], random-forest learning [3], and support vector machine (SVM) [4] models. In this stage, more studies began to focus on STPFF with the gradual development of URT. For example, Roos *et al.* combined dynamic Bayesian networks with Gaussian mixture models to conduct STPFF in URT [5]. Li *et al.* [6] built multiscale radial-basis-function networks to perform STPFF in URT. Some studies combined ARIMA with wavelet decomposition, SVM and BPNNs to conduct STPFF in URT [7-9]. These hybrid models exhibit significantly better prediction precision than most mathematical statistics-based models. However, most of these models cannot consider spatial correlations in the model formulation. Furthermore, researchers always take one or several subway stations as examples to verify their models. For simultaneous forecast in a network with several hundred subway stations, such models cannot perform well.

As a branch of machine learning, deep learning models become prevailing nowadays. Since deep-belief networks were initially proposed in 2006 [10], deep neural networks (DNNs) have received enormous attention for applications in computer science. With the rapid expansion of traffic infrastructures, numerous researchers have applied DNNs to achieve STPFF in URT and road traffic because of their powerful ability to capture spatiotemporal, topological, and much other

Manuscript received July 18, 2019; date of current version May 8, 2019. This work was supported by the National Natural Science Foundation of China (grant numbers 71871027 and 51978044).

Jinlei Zhang, Feng Chen, and Yadi Zhu are with the School of Civil Engineering, Beijing Jiaotong University, No.3 Shangyuancun, Haidian District, Beijing 100044, China. Feng Chen is with Chang'an University, Middle-section of Nan'er Huan Road, Xi'an 710064, China. Feng Chen is also with Beijing Engineering and Technology Research Centre of Rail Transit Line Safety and Disaster Prevention, No.3 Shangyuancun, Haidian District, Beijing 100044, China (e-mail: 17115275@bjtu.edu.cn; fengchen@bjtu.edu.cn; yadizhu@bjtu.edu.cn). *(Corresponding author: Feng Chen.)*
Zhiyong Cui is with the Department of Civil and Environmental Engineering, University of Washington, Seattle, WA 98195, USA. Yinan Guo is with the Information School, University of Washington, Seattle, WA 98105, USA (e-mail: zhiyongc@uw.edu; yguo17@uw.edu).

information.

The typical deep learning models include long short-term memory (LSTM) [11] and gated recurrent units (GRUs) [12]. Although Guo *et al.* [13] proposed a hybrid SVM–LSTM model to predict short-term abnormal flow in URT, they did not take spatial correlations between stations into account. Tang *et al.* [14] proposed an LSTM model to forecast outflows, wherein time-cost and spatial-correlation matrices were used in this study. Li *et al*. [15] built a dynamic radial basis function (RBF) neural network to forecast outflows. These two studies only involved outflow, but inflow is more significant for URT operation. Furthermore, they did not consider spatial correlations.

Some models are also prevailing, such as convolutional neural networks (CNNs) [16], convolutional LSTM (ConvLSTM) [17], and stacked autoencoders (SAEs) [18]. Although Liu *et al.* [19] built a deep learning architecture by considering spatiotemporal, environmental, and operational factors, their model could be applied only to several subway stations rather than a whole network. Since residual neural networks (ResNets) have been proposed [20], researchers have applied them to road-traffic scenarios such as bus [21] and taxi [22] flow predictions. To the best of our knowledge, however, there have been fewer such applications of ResNets to STPFF in URT. Li *et al.* [23] introduced an innovative deep belief network to conduct multi-step predictions. The missing values were estimated during data processing [24]. Overall, these models did not consider the topological information of the network.

Although many models have considered spatiotemporal and topological information [25-28], external factors such as weather conditions and air quality were often neglected. Since passengers may adjust their trips when encountering bad weather conditions or heavily polluted atmosphere, such external factors are also important in traffic prediction. Li *et al.* [29] introduced a graph CNN to conduct STPFF in URT. Similarly, Han *et al.* [30] built spatiotemporal graph CNNs to predict short-term ridership in a citywide metro network. Although these studies extracted spatial correlations between stations from network topology, they did not consider external factors, such as weather conditions, events, or air quality.

In this study, we build a deep learning architecture called "ResLSTM," combing ResNet, GCN, and attention LSTM to conduct STPFF in URT on a network scale. In addition to the spatiotemporal correlations between subway stations, the topological relationships between them, as well as weather conditions and air quality, are also incorporated into ResLSTM to determine how such factors affect passenger travel. We compare the performance of the proposed ResLSTM model with that of several prevailing state-of-the-art models. The experimental results show that the proposed ResLSTM model outperforms those baseline models. The main contributions of the proposed architecture are as follows.

1) The proposed ResLSTM model considers not only spatiotemporal features but also network topology as well as weather conditions and air quality. Moreover, the ability to make real-time predictions with high precision on a network scale is realized.
2) The proposed ResLSTM architecture is so robust that there is only a negligible effect on prediction precision when deleting one of the four branches that constitute the architecture.
3) The influences of weather conditions and air quality on prediction accuracy are quantified. The prediction accuracy is improved in terms of the evaluation metrics.

The remainder of this paper is organized as follows. In Section II, methodologies of ResNet, GCN, and attention LSTM are presented. In Section III, the architecture of the proposed ResLSTM model is described. In Section IV, case study results are discussed. The main findings and limitations of the current study and their significance are summarized and directions for future research are proposed in Section V.

## II. METHODOLOGY

The architecture of our model was primarily designed based on ResNet, GCN, and attention LSTM. Therefore, we briefly introduce the respective methodologies of these components in this section.

### A. ResNet

Network passenger flows can be treated as preprocessed images when applying CNN [16]. Previous studies have shown that deeper models can extract more enriched features [31]. However, deeper models are not always better because of vanishing or exploding gradients [32]. Hence, in 2015 He *et al.* [20] proposed ResNet, which contains a skip connection, as shown in Fig. 1 (a). The purpose of the ResNet model is to train network output as follows:

$$X_{l+1} = F(X_l) + X_l, \qquad (1)$$

where $X_l$ and $X_{l+1}$ represent residual block input and output, respectively.

In this study, we adopt an improved residual block [33], as shown in Fig. 1 (b). In the improved residual block, gradients can be unimpededly passed to any earlier layers through shortcut connections, thus solving the vanishing or exploding gradient problem. An example of a residual block with 32 filters used in our study is shown in Fig. 2, where "Conv" indicates a convolutional layer, "BN" denotes a batch-normalization layer, and "ReLU" represents an activation layer.

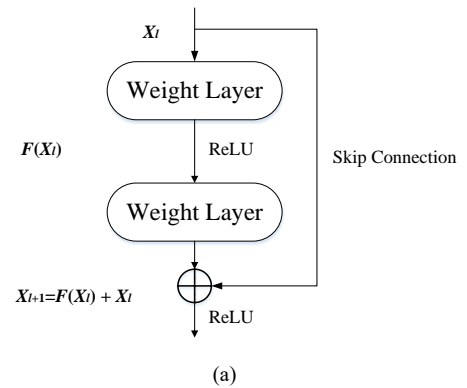

(a)

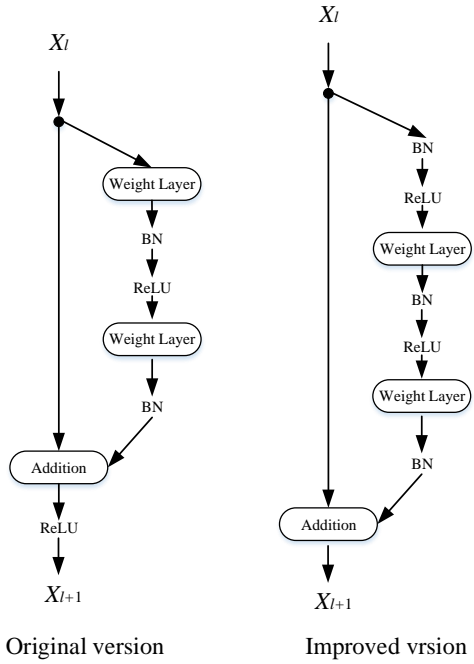

Original version     Improved vrsion

(b)

Fig. 1. Residual block (a) and comparison of original and improved residual blocks (b).

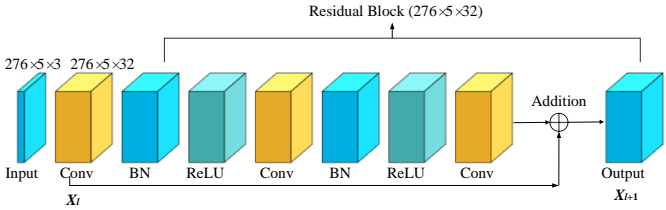

Fig. 2. Dimensions of tensors flowing in the residual block.

### B. GCN

CNN-related models generally treated traffic networks as grid matrices. However, such processes neglect the influence of network topology on prediction precision. Therefore, we applied GCN, as shown in Fig. 3, to capture URT-network topological dependencies. Take general graph $G = (V, E)$ as an example, where $V$ is the set of vertices and $E$ is the set of edges representing the relationships between adjacent nodes, the GCN functions can then be defined as follows [34]:

$$H^{l+1} = f(H^l, A) = \sigma(\widehat{D}^{-\frac{1}{2}}\hat{A}\widehat{D}^{-\frac{1}{2}}H^l W^l), \tag{2}$$

$$H^{l'} = \widehat{D}^{-\frac{1}{2}}\hat{A}\widehat{D}^{-\frac{1}{2}}H^l, \tag{3}$$

where $\hat{A} = A + I$, $A \in R^{n \times n}$ is the adjacency matrix, $I$ is the identity matrix, $\widehat{D}$ is the diagonal node-degree matrix of $\hat{A}$, $W$ is the weight matrix of the $l^{th}$ layer, $H \in R^{n \times m}$ is the feature matrix, wherein $m$ is the number of features of each of the $n$ nodes, $H' \in R^{n \times m}$ is the feature matrix with topological information, and $\sigma(\cdot)$ is an activation function.

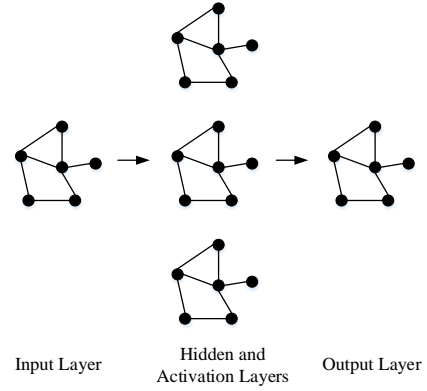

Fig. 3. Graph convolutional network.

However, some previous works have shown that stacking multiple GCN layers not only leads to higher complexity during backpropagation but also to gradient vanishing [35, 36], thereby degrading the performance of deeper GCNs. Moreover, oversmoothing, which means multiple features of the same vertex converge to the same value, is also a common problem that arises with deeper GCNs [37]. Therefore, in this study, we extend GCN into ResNet GCN to mitigate such drawbacks. Each input matrix $In$ is treated as a graph signal, which is then transformed according to

$$In' = \widehat{D}^{-\frac{1}{2}}\hat{A}\widehat{D}^{-\frac{1}{2}}In, \tag{4}$$

where $\widehat{D}^{-\frac{1}{2}}\hat{A}\widehat{D}^{-\frac{1}{2}}$ is the symmetric normalized Laplacian, as shown in (3), $In \in R^{s \times t}$ is the input, where $s$ is the number of subway stations and $t$ represents historical timesteps for each station. The transformed $In'$ has the same shape as $In$ and contains rich network topological information that was subsequently used as ResNet input.

### C. Attention LSTM

The LSTM attention mechanism has been shown to be effective in predicting traffic flow. Because the attention mechanism has been successfully used for machine translation since its introduction [38], many researchers have applied it to STPFF [39-41]. Therefore, to capture the different weights of features extracted from former network layers, we introduced attention LSTM to our model.

Conventional attention LSTM is used to capture the weight scores of different timesteps, usually by assigning heavier weight scores to adjacent timesteps and lower ones to those further apart. However, traffic prediction models, which are affected by many factors such as weather conditions, passenger enter and exit flows, and network topology, are so sophisticated that assigning weight scores based on recentness is insufficient [41]. Therefore, based on previous work by Wu *et al.*[41], we use a fully connected network to obtain weights that can be scored according to the input or LSTM output. Preliminary test results indicated that the latter was more effective; therefore, LSTM output weight is automatically scored in the proposed model. Let matrix $Out \in R^{m \times n}$ be the LSTM output, where $m$ and $n$ represent the timesteps and number of features of each timestep, respectively. Then, the attention-based output ($Out'$) can be obtained by

$$A = f(W \circ Out + b), \quad (5)$$
$$Out' = A \circ Out, \quad (6)$$

where $A$ is a weight matrix whose shape is identical to that of $Out$, "$\circ$" denotes the Hadamard product, $f$ represents the fully connected layer (which can be activated by different activation functions such as sigmoid functions), $W$ is the weight matrix of $f$, and $b$ is the bias.

## III. MODEL DEVELOPMENT

Herein, we describe the ResLSTM model architecture, as shown in Fig. 4, which comprises four branches. All input data are obtained for times $t - n$ to $t$, and output data are inflows from time $t + 1$. Branch 1 uses the inflows to capture spatiotemporal features. Branch 2 is identical to Branch 1 except it uses the outflows. Branch 3 extracts network topological information. Branch 4 represents the impacts of weather conditions and air quality on prediction precision. Moreover, attention LSTM is used in the trunk to obtain the output data. A detailed model architecture description is presented in subsections A to E.

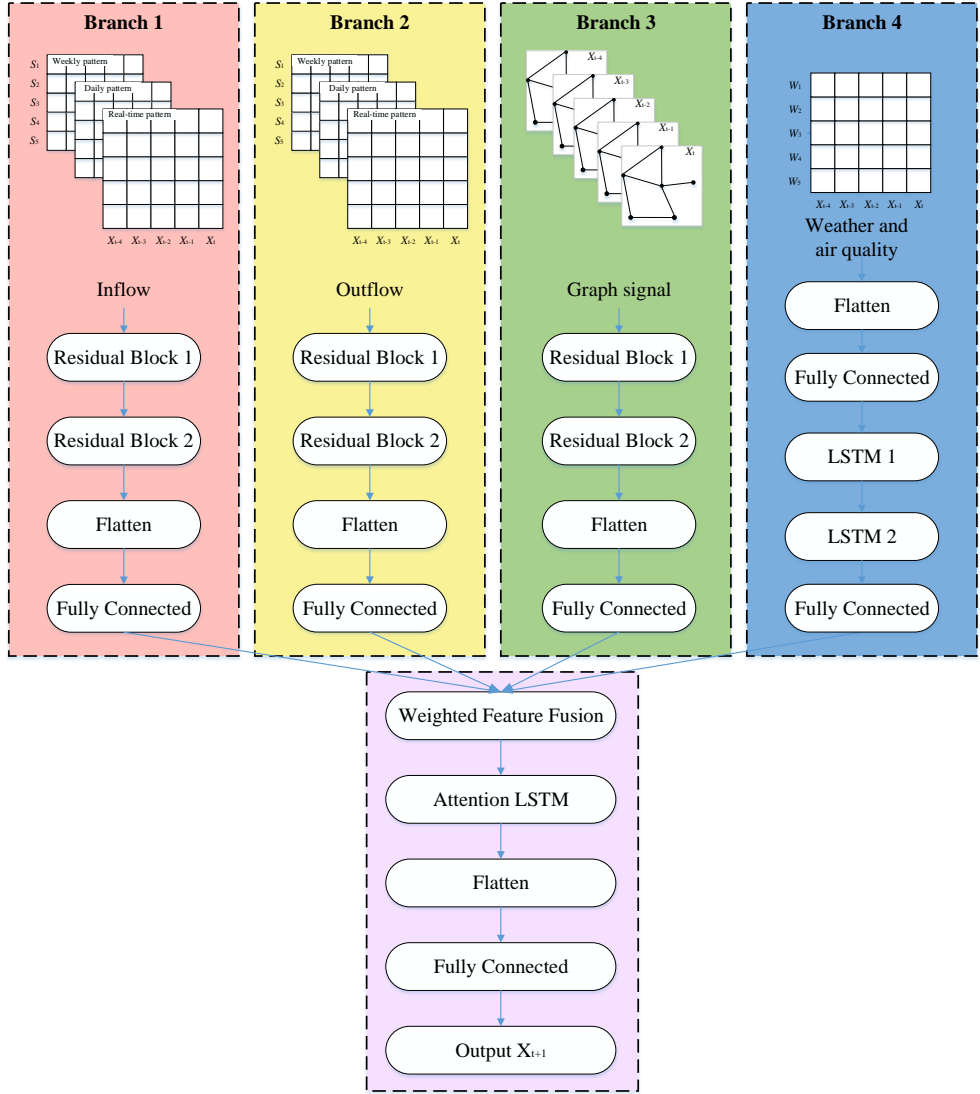

Fig. 4. ResLSTM model architecture.

### A. Branch 1: Inflow

Prior knowledge of historical inflow is the most important for predicting network inflow. Therefore, Branch 1 captures the inflow. In previous studies, the relationship between inflow and outflow has always been indicated by a model showing two channels [22, 29, 30]: one corresponding to inflow and the other to outflow. However, when three patterns (such as real-time, daily, and weekly patterns) must be considered, the model should have three branches, which significantly increases its complexity. Therefore, we propose the method of separately considering inflow and outflow. Initially, for a given station in a subway-station network, inflow and outflow are only slightly related, which is completely different from a road-traffic network; a subway station showing large inflow may show only small outflow. Furthermore, treating inflow and outflow separately can decrease model complexity without decreasing prediction precision.

Nowadays, real-time inflow and outflow in a subway station can be obtained by automatic fare collection (AFC) systems. Therefore, we considered three inflow patterns: real-time, daily,

and weekly patterns. The inflow time series is given by

$$X_{s,t}^p = \begin{pmatrix} x_{1,t-n}^p & x_{1,t-n+1}^p & x_{1,t-n+2}^p & \cdots & x_{1,t}^p \\ x_{2,t-n}^p & x_{2,t-n+1}^p & x_{2,t-n+2}^p & \cdots & x_{2,t}^p \\ x_{3,t-n}^p & x_{3,t-n+1}^p & x_{3,t-n+2}^p & \cdots & x_{3,t}^p \\ \vdots & \vdots & \vdots & \ddots & \vdots \\ x_{s,t-n}^p & x_{s,t-n+1}^p & x_{s,t-n+2}^p & \cdots & x_{s,t}^p \end{pmatrix}, \quad (7)$$

where $s$ is the number of subway stations and $t$ represents the historical timesteps for each station. Stations are ordered in columns according to their line number, e.g., line 1, line 2, etc. In each line, adjacent stations are placed in adjacent rows according to train direction, where "$p$" represents different patterns. If $p$ represents real-time, daily, or weekly patterns, $X_{s,t}^r$, $X_{s,t}^d$, and $X_{s,t}^w$ represent inflow time series corresponding to the same day, previous day, or previous week, respectively.

As shown in Branch 1, to predict inflow for time $t + 1$, we organize time series from three patterns for times $t$-$4$ to $t$ into a single three-channel "image." The Branch 1 input is given by

$$I_1 = (X_{s,t}^r, X_{s,t}^d, X_{s,t}^w). \quad (8)$$

Data are input into two residual blocks, the first showing 32 filters and the second showing 64. Then, the data are flattened and fully connected with 276 neurons. The Branch 1 output data are then input into the feature-fusion section.

*B. Branch 2: Outflow*

Outflow processing for Branch 2 is identical to the inflow processing for Branch 1. Hence, the Branch 2 input is given by

$$I_2 = (X'^{r}_{s,t}, X'^{d}_{s,t}, X'^{w}_{s,t}), \quad (9)$$

where $X'$ represents the outflow.

*C. Branch 3: Graph signal*

Traffic-network topology has been proven to be important for STPFF [29, 30]. To overcome the drawbacks discussed in Section II-B, we use a ResNet GCN, as shown in Branch 3 in Fig. 4, to capture the influence of network topology. We only consider the real-time pattern because the network topology does not change. According to (4), (7), and (8), the original input for ResNet GCN is

$$I_3 = \hat{D}^{-\frac{1}{2}} \hat{A} \hat{D}^{-\frac{1}{2}} (X_{s,t}^r). \quad (10)$$

Input data are subsequently processed according to the method described for Branch 1.

*D. Branch 4: Weather conditions and air quality*

Although a few researchers have considered the impact of weather conditions on STPFF [19], to the best of our knowledge, none have considered the impact of air quality on STPFF. However, weather conditions and air quality are both essential information for people scheduling their travel plans. For example, cold weather and heavily polluted air often impede passengers' nonemergency, nonessential travels.

The dataset used in this branch, therefore, contained two subsets: weather conditions showing the real-time temperature (℃), dew point temperature (℃), relative humidity (%), and wind speed (m/s), all of which were recorded every half-hour, as shown in TABLE I; and air quality showing the real-time air quality index (AQI) and concentrations of atmospheric particulate matter (PM2.5 and PM10), $SO_2$, $NO_2$, CO, and $O_3$ (μg/m³), all of which were recorded every hour, as shown in TABLE II.

TABLE I

EXAMPLES OF WEATHER-CONDITION DATA

| Date/Time | Temperature (℃) | Dew point temperature (℃) | Relative humidity (%) | Wind speed (m/s) |
|---|---|---|---|---|
| 2016/2/29 5:00 | -6 | -17 | 42 | 4 |
| 2016/2/29 5:30 | -6 | -17 | 42 | 4 |
| 2016/2/29 6:00 | -5 | -17 | 39 | 11 |
| 2016/2/29 6:30 | -5 | -18 | 36 | 7 |
| 2016/2/29 7:00 | -6 | -18 | 39 | 4 |
| 2016/2/29 7:30 | -6 | -16 | 46 | 7 |
| 2016/2/29 8:00 | -3 | -16 | 37 | 4 |

TABLE II

EXAMPLES OF AIR-QUALITY DATA

| Date/Time | AQI | PM2.5 | PM10 | $SO_2$ | $NO_2$ | CO | $O_3$ |
|---|---|---|---|---|---|---|---|
| 2016/2/29 5:00 | 18 | 11 | 16 | 5 | 36 | 0.5 | 38 |
| 2016/2/29 6:00 | 21 | 13 | 21 | 6 | 40 | 0.5 | 37 |
| 2016/2/29 7:00 | 20 | 14 | 20 | 5 | 38 | 0.5 | 40 |
| 2016/2/29 8:00 | 20 | 12 | 20 | 5 | 37 | 0.5 | 41 |
| 2016/2/29 9:00 | 24 | 15 | 24 | 6 | 35 | 0.5 | 47 |
| 2016/2/29 10:00 | 25 | 17 | 24 | 7 | 33 | 0.6 | 53 |
| 2016/2/29 11:00 | 28 | 19 | 26 | 7 | 33 | 0.6 | 54 |

If we conduct STPFF at time granularity (TG) = 10 min, the weather-condition data from 05:00 to 05:10 will share the recorded data from 05:00 to 05:30, as shown in the first row of TABLE I. Similarly, the corresponding air-quality data from 05:00 to 05:10 will share the recorded data from 05:00 to 06:00, as shown in the first row of TABLE II. We obtain the preprocessed input data for Branch 4 as

$$I_4 = X_{w,t} = \begin{pmatrix} x_{1,t-n} & x_{1,t-n+1} & x_{1,t-n+2} & \cdots & x_{1,t} \\ x_{2,t-n} & x_{2,t-n+1} & x_{2,t-n+2} & \cdots & x_{2,t} \\ x_{3,t-n} & x_{3,t-n+1} & x_{3,t-n+2} & \cdots & x_{3,t} \\ \vdots & \vdots & \vdots & \ddots & \vdots \\ x_{w,t-n} & x_{w,t-n+1} & x_{w,t-n+2} & \cdots & x_{w,t} \end{pmatrix}, \quad (11)$$

where $w$ represents the 11 indicators used for weather-condition and air-quality data.

Preprocessed input data are flattened and subsequently added to the fully connected layer to obtain weighted indicators. Stacked LSTM with 128 and 276 neurons for the first and second layers, respectively, is then applied. The output data are then input into the feature-fusion section.

*E. Feature fusion*

Because the data output from the four branches is identical in shape, weighted feature-fusion is easily implemented according

to

$$\text{Fusion} = W_1 \circ O_1 + W_2 \circ O_2 + W_3 \circ O_3 + W_4 \circ O_4, \quad (12)$$

where $O_1$, $O_2$, $O_3$, and $O_4$ are outputs from the four branches, $W$ is the corresponding weight vector used to capture degrees of impact of different features, and "∘" represents the Hadamard product. The $W$ has the same shape dimensions with the outputs. The values in the weight vector $W$ are initialized randomly before training and can be updated during backpropagation.

Attention LSTM, described in Section II-C, is then applied after feature fusion [14]. The LSTM output is subsequently flattened and fully connected with 276 neurons to generate the final output.

## IV. CASE STUDY RESULTS AND DISCUSSION

Herein, we describe the dataset, present the detailed model configuration, and compare the performances of the different models evaluated.

### A. Dataset description

The AFC data used in this study were collected from the Beijing subway between 05:00 and 23:00 for five consecutive weeks from February 29 to April 3, 2016. There were 17 lines and 276 subway stations (excluding the airport express line and the stations on it) in March 2016 in Beijing. Only data from the 25 workdays in the target period are applied to this study, constituting 130 million records. Every record contains the card number, entry-station number, exit-station number, entry time, exit time, entry-station name, and exit-station name. Inflow and outflow time series are extracted according to (7). The TGs used in our study are 10, 15, and 30 min. Therefore, it is easy to integrate the results predicted for TG = 10 and 15 min into those predicted for TG = 30 min to compare prediction performances for different TGs. Examples of weather-condition and air-quality datasets are provided in Section III-D.

### B. Model configuration

The model was implemented using TensorFlow and Keras [42]. We used data from the first four weeks to train the models and data from the final week to test them. The validation split rate was set as 0.2 to calibrate the model. To balance the tradeoff between model training time and prediction precision, we used the previous five network timesteps to forecast the next one by trial and error. For Branch 1, the first residual block has 32 filters; the second has 64. The kernel size is 3*3. The fully connected layer consists of 276 neurons. Branches 1, 2, and 3 show the same configuration. For Branch 4, the fully connected layers consist of 276 neurons and the two LSTM layers consist of 128 and 276 neurons. For feature fusion, the attention LSTM and final fully connected layers consist of 128 and 276 neurons, respectively.

To avoid inappropriate parameter initialization, we trained the proposed ResLSTM and baseline models for multiple times before the hyperparameters are determined. During the training procedure, we used the Model Checkpoint and Early Stopping technique to save the best model and avoid overfitting [42].

Before Early Stopping, both the training loss and validation loss are shown in Fig. 5. As is shown, the training loss and validation loss present a significant vibration for the first 150 epochs. After the 150 epochs, the two losses remain stable and only suffer from a slight vibration, which shows the strong robustness of the proposed model.

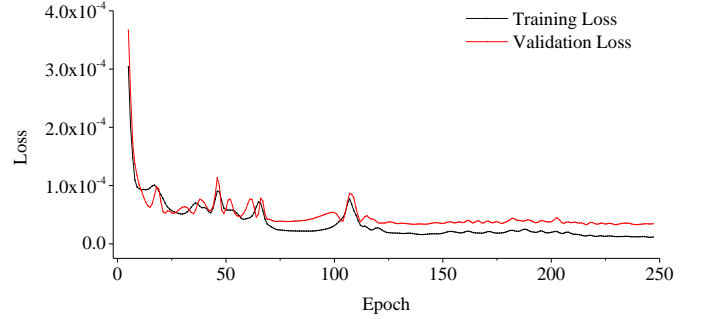

Fig. 5. Variation for training loss and validation loss

### C. Baseline models

In this study, we compare the performances of several models. Note that when using ARIMA to conduct STPFF on a subway network, we had to build 276 models representing each of the 276 stations. Except for ARIMA, all the benchmark models were used to obtain overall results for the 276 stations by training only a single model, and their optimizers are Adam with a learning rate of 0.0001. For the five variants of ResLSTM, the other configurations are the same with ResLSTM. The specific configurations are as follows.

**ARIMA** [43]**:** A representative conventional mathematical statistics-based model. We used Expert Modeler in the Statistical-Package-for-the-Social-Sciences (SPSS®) software (IBM Corp., USA) to obtain the best ARIMA results automatically.

**BPNN:** The BPNN has two hidden layers, each containing 100 neurons.

**Support Vector Regression (SVR)** [44]**:** The kernel of SVR in scikit learn is set as radial-basis function (RBF–SVR). The regularization parameter C is set as 1.0. The tolerance for stopping criterion is set as 0.001.

**Vanilla RNN, LSTM** [11]**, and GRU:** All of them have two kernel layers, each containing 100 neurons.

**CNN and ConvLSTM** [16, 45]**:** Both of them have two kernel layers with 32 and 64 filters, respectively. The kernel size is 3*3.

**ResLSTM-GCN:** We only adopted Branch 3.

**ResLSTM-No graph:** We deleted branches 3.

**ResLSTM-No W&A:** We deleted branches 4.

**ResLSTM-No A:** We deleted the air quality data.

**ResLSTM-TC:** We organized the inflow and outflow as two channels. That is, Branches 1 and 2 are transformed into three branches, each containing one pattern and two channels for inflow and outflow.

### D. Loss function and evaluation metrics

We adopt end-to-end training to optimize the model. The mean-squared error (MSE) is used as the loss function. The optimizer is "Adam" with a learning rate of 0.001. We apply three indicators to evaluate model performance: root-mean-squared error (RMSE), mean-absolute error (MAE), and weighted-mean-absolute-percentage error (WMAPE). They are given by:

$$Loss = \text{MSE} = \frac{1}{n}\sum_{i=1}^{n}(y_i - \hat{y}_i)^2, \quad (13)$$

$$RMSE = \sqrt{\frac{1}{n}\sum_{i=1}^{n}(y_i - \hat{y}_i)^2}, \quad (14)$$

$$MAE = \frac{1}{n}\sum_{i=1}^{n}|(y_i - \hat{y}_i)|, \quad (15)$$

$$WMAPE = \sum_{i=1}^{n}\left(\frac{y_i}{\sum_{j=1}^{n}y_j}\left|\frac{y_i-\hat{y}_i}{y_i}\right|\right), \quad (16)$$

where $y_i$ is the actual value, $\hat{y}_i$ is the predicted value, $n$ is the number of samples, and $\sum_{j=1}^{n}y_j$ is the sum of the actual values.

*E. Results and discussion*

*1) Network-wide prediction performance*

The prediction performances are shown in TABLE III and Fig. 5. As shown in TABLE III, deep learning models considerably outperform mathematical statistics-based and machine-learning-based models in most cases. The RBF–SVR model performs the worst, even worse than ARIMA. The reason may be that we built 276 models representing each of the 276 stations when conducting ARIMA, while the SVR was used to obtain overall results for the 276 stations by training only a single model. Moreover, the SVR is unsuitable for the regression of large datasets due to its higher computation cost. The second-worst model is ARIMA. Although each subway station has its own individual model in ARIMA, the prediction performance is poor because ARIMA cannot capture the comprehensive nonlinear features of passenger flows.

Among deep learning models, all the convolution-based models perform better than the recurrence-based models when only a single model is used for network predictions. Both LSTM and GRU perform better than Vanilla RNN, as expected. ConvLSTM performs better than CNN because ConvLSTM can capture more temporal information. However, the performance of ConvLSTM worsens with increasing TG. A decrease in the number of samples may account for such phenomena.

Among ResLSTM and its five variants, the complete ResLSTM performs best because many features including inflow, outflow, network topology, as well as weather conditions and air quality are fully taken into account. It is worth mentioning that the proposed architecture presents strong robustness; that is, even when one branch is deleted, prediction results do not change significantly (see results of ResLSTM-No Graph, ResLSTM-No W&A, and ResLSTM-No A).

For the topological information, satisfactory results are obtained using only Branch 3 (see results of ResLSTM-GCN), which strongly demonstrates the robustness of the proposed architecture. Although all the models show similar prediction performances when TG = 10 min, ResLSTM begins to show its superior prediction performance and the gaps between the performances of ResLSTM and its variants widen when TG is increased from 10 to 30 min. Moreover, ResLSTM always performs best regardless of whether TG = 10, 15, or 30 min. Comparing the prediction performances of ResLSTM-No Graph and ResLSTM shows that network topology has some influence on prediction precision.

For the weather conditions and air quality, comparing the prediction performances of ResLSTM-No W&A, ResLSTM-No A, and ResLSTM, we can infer using common sense that the introduction of weather-condition and air-quality datasets increases prediction precision. If the weather is very cold or air pollution is critically high (e.g., higher PM2.5 and PM10), people will reduce or eliminate unnecessary travel, meaning these external factors may affect passenger volume. Moreover, this influence is quantified, with the RMSE, MAE, and WMAPE decreasing from 60.13 to 56.96, 34.14 to 32.58, and 6.43 to 6.13%, respectively, by considering weather and air quality when TG = 30 min.

Comparing the prediction performances of ResLSTM-TC and ResLSTM, we can infer that treating passenger flow separately will not only save computation cost but will also retain prediction precision.

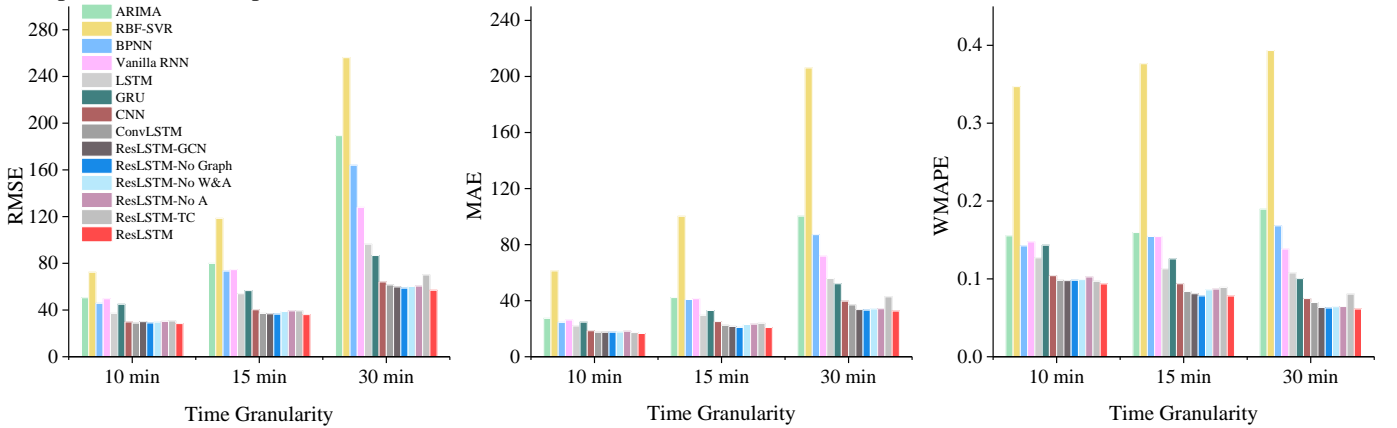

Fig. 6. Comparison of prediction performances obtained for different models and TGs.

TABLE III
COMPARISON OF PREDICTION PERFORMANCES OBTAINED USING DIFFERENT TGS IN DIFFERENT MODELS

| Category | TG Indicators | 10 min RMSE | 10 min MAE | 10 min WMAPE | 15 min RMSE | 15 min MAE | 15 min WMAPE | 30 min RMSE | 30 min MAE | 30 min WMAPE |
|---|---|---|---|---|---|---|---|---|---|---|
| MS | ARIMA | 50.5436 | 27.3968 | 15.53% | 79.9580 | 42.2139 | 15.95% | 189.3329 | 100.3590 | 18.95% |
| ML | RBF-SVR | 72.3753 | 61.2418 | 34.70% | 118.5395 | 100.2861 | 37.63% | 256.0734 | 206.0826 | 39.33% |
| ML | BPNN | 45.6946 | 24.5322 | 14.24% | 73.3083 | 40.8457 | 15.45% | 163.9720 | 87.1259 | 16.81% |
| DL | Vanilla RNN | 49.6506 | 26.2275 | 14.75% | 74.6021 | 41.4159 | 15.43% | 127.9259 | 71.8328 | 13.84% |
| DL | LSTM | 37.1903 | 21.9925 | 12.71% | 53.9216 | 29.5340 | 11.29% | 96.3534 | 55.8265 | 10.76% |
| DL | GRU | 44.9115 | 24.7353 | 14.33% | 56.6550 | 33.0309 | 12.56% | 86.6984 | 52.0717 | 10.03% |
| DL | CNN | 29.8125 | 18.5460 | 10.413% | 40.2673 | 25.1231 | 9.375% | 64.0458 | 39.6867 | 7.472% |
| DL | ConvLSTM | 28.7943 | 17.4780 | 9.814% | 37.0923 | 22.4236 | 8.380% | 61.4978 | 36.9768 | 6.962% |
| DL | ResLSTM-GCN | 30.0706 | 17.4331 | 9.775% | 36.7801 | 21.6615 | 8.099% | 59.7289 | 33.6304 | 6.334% |
| DL | ResLSTM-No Graph | 28.9342 | 17.4870 | 9.827% | 36.4708 | 20.9008 | 7.815% | 58.7114 | 33.2383 | 6.260% |
| DL | ResLSTM-No W&A | 29.6219 | 17.6344 | 9.911% | 38.7386 | 23.0008 | 8.589% | 60.1340 | 34.1360 | 6.428% |
| DL | ResLSTM-No A | 30.3012 | 18.2160 | 10.244% | 39.3208 | 23.2970 | 8.699% | 60.5372 | 34.2644 | 6.451% |
| DL | ResLSTM-TC | 30.6557 | 17.2621 | 9.692% | 39.2743 | 23.8779 | 8.901% | 70.0569 | 42.7349 | 8.046% |
| DL | **ResLSTM** | **28.3661** | **16.6318** | **9.352%** | **36.0444** | **20.8783** | **7.805%** | **56.9649** | **32.5819** | **6.134%** |

Note: MS indicates a mathematical statistics-based model. ML indicates a machine-learning-based model. DL indicates a deep learning model.

*2) Prediction performance of individual stations*

We chose three typical stations to analyze the prediction performance of individual stations. The first one is Tiantongyuan station, a large community station with millions of people living around it. The second one is Xizhimen station, a typical traffic hub with three subway lines interchanging there and many bus stops nearby. The last one is Beijing West Railway station, a subway station near the large railway station. The predicted results for these three typical stations in three time granularities are shown as follows.

For Tiantongyuan station, as shown in Fig. 7, the predicted values are always in line with the actual values for all the three time granularities, no matter in peak periods or non-peak periods, showing the strong robustness of the ResLSTM. Because Tiantongyuan station is located around a large residential community, there are thousands of people commuting in the morning. Therefore, the morning peak is obvious and the flow regularity is significant, which contributes to the performance.

As a large traffic hub, the inflows from Xizhimen station presents double-peak characteristics. As shown in Fig. 8, the model also keeps a favorable performance for all the three time granularities, especially in peak periods.

Different from the two stations, the inflows from Beijing West Railway station represent low regularities and suffer a significant variation. However, as shown in Fig. 9, the model can still capture the variation trend. Moreover, with the time granularities increasing from 10 min to 30 min, the fitting becomes better, indicating that the prediction performance is improved.

In summary, the proposed model can conduct a precise prediction not only on a network scale but also on a station scale.

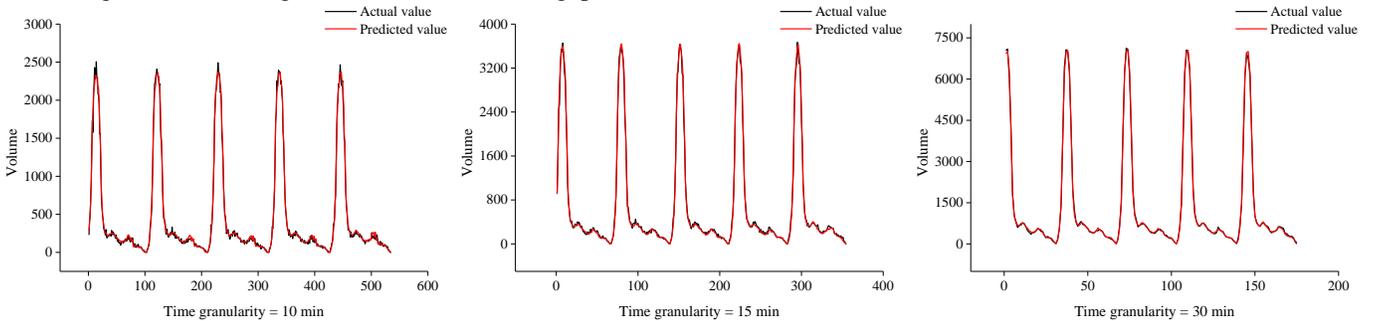

Fig. 7 Comparison of actual values and predicted values for Tiantongyuan Station

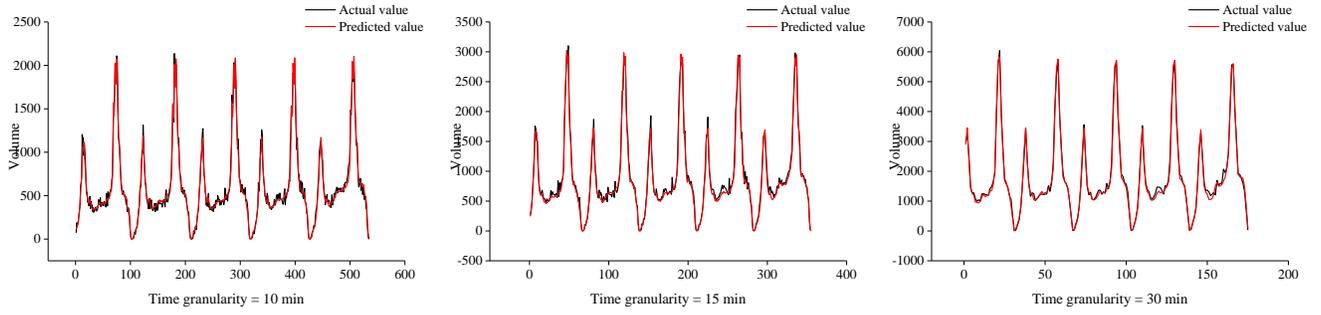

Fig. 8 Comparison of actual values and predicted values for Xizhimen Station

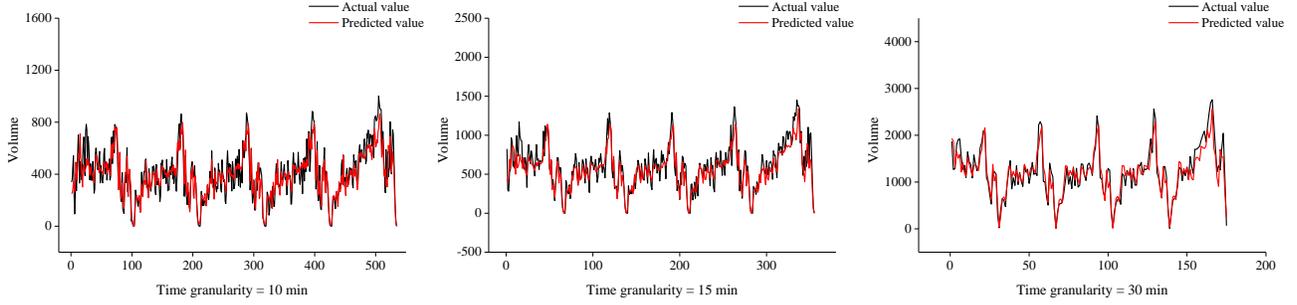

Fig. 9 Comparison of actual values and predicted values for Beijing West Railway Station

*3) Prediction performance in different TGs*

To compare the prediction precisions obtained for different TGs, we aggregated the results obtained when TG = 10 and 15 min into corresponding results obtained when TG = 30 min and then computed corresponding evaluation indicators. The results are shown in TABLE IV, in which it can be observed that prediction precision gradually increases with increasing TG. When TG is 10*3 min and 30 min, the RMSE, MAE, and WMAPE all decrease from 61.55 to 56.96, 35.22 to 32.58, and 6.63 to 6.13%, respectively. From a statistical perspective, it is because the passenger-flow similarity and regularity increase when flows are aggregated under larger TGs, thereby contributing to better overall prediction precision.

In summary, the proposed ResLSTM model presents a satisfactory ability to conduct STPFF in URT. It also shows strong robustness that is favorable for practical real-world applications.

TABLE IV
COMPARISON OF PREDICTION PRECISIONS OBTAINED USING DIFFERENT TGS

|  | RMSE | MAE | WMAPE |
| --- | --- | --- | --- |
| 10*3 = 30 min | 61.5458 | 35.2239 | 6.633% |
| 15*2 = 30 min | 58.0825 | 33.2736 | 6.265% |
| 30 min | 56.9649 | 32.5819 | 6.134% |

## V. CONCLUSION

In this study, we proposed a deep learning architecture called ResLSTM incorporating ResNet, GCN, and attention LSTM. To the best of our knowledge, this is the first time that air-quality indicators such as PM2.5 and PM10 have been considered in STPFF. Moreover, the influences of weather conditions and air quality on prediction precision were quantified. In addition, we combined GCN and ResNet to overcome the drawbacks of GCN. The main conclusions are summarized as follows.

1) The ResLSTM model was able to capture not only the spatiotemporal features of passenger flows but also network topological information as well as the influences of weather conditions and air quality on prediction precision.
2) The ResLSTM model showed strong robustness, which is essential for real-world applications. Moreover, the prediction precisions were favorable when conducting STPFF on a network scale for nearly 300 subway stations.
3) Weather conditions and air quality were proven to have considerable influence on prediction precision and the influence was quantified.
4) The prediction precision increased with increasing TG because of the higher similarity and regularity when passenger flows were aggregated under a higher TG.

However, there are several limitations to our study. For example, we did not consider weekend passenger flows owing to substantial fluctuations and less regularity. Multistep predictions should also be explored in the future. Moreover, model interpretability is currently poor because the proposed model is a "black box", wherein data are fed to obtain satisfactory predictions without disclosing the applied process. Future studies should attempt to compensate for these limitations.

## REFERENCES

[1] E. I. Vlahogianni, J. C. Golias and M. G. Karlaftis, "Short-term traffic forecasting: Overview of objectives and methods," Transport reviews, vol. 24, pp. 533-557, 2004.
[2] P. Wang and Y. Liu, "Network traffic prediction based on improved BP wavelet neural network," in 2008 4th International Conference on

[3] N. Zarei, M. A. Ghayour and S. Hashemi, "Road traffic prediction using context-aware random forest based on volatility nature of traffic flows," in Asian Conference on Intelligent Information and Database Systems, 2013, pp. 196-205.

[4] H. Su and S. Yu, "Hybrid GA based online support vector machine model for short-term traffic flow forecasting," in International Workshop on Advanced Parallel Processing Technologies, 2007, pp. 743-752.

[5] J. Roos, S. Bonnevay and G. Gavin, "Dynamic Bayesian networks with Gaussian mixture models for short-term passenger flow forecasting," in International Conference on Intelligent Systems and Knowledge Engineering, 2017, pp. 1-8.

[6] Y. Li, X. Wang, S. Sun, X. Ma, and G. Lu, "Forecasting short-term subway passenger flow under special events scenarios using multiscale radial basis function networks," Transportation Research Part C: Emerging Technologies, vol. 77, pp. 306-328, 2017.

[7] X. Wang, N. Zhang, Y. Chen, and Y. Zhang, "Short-term forecasting of urban rail transit ridership based on ARIMA and wavelet decomposition," in AIP Conference Proceedings, 2018, pp. 040025.

[8] X. Wang, N. Zhang, Y. Zhang, and Z. Shi, "Forecasting of Short-Term Metro Ridership with Support Vector Machine Online Model," Journal of Advanced Transportation, vol. 2018, 2018.

[9] L. Li, Y. Wang, G. Zhong, J. Zhang, and B. Ran, "Short-to-medium term passenger flow forecasting for metro stations using a hybrid model," KSCE Journal of Civil Engineering, vol. 22, pp. 1937-1945, 2018.

[10] G. E. Hinton, S. Osindero and Y. Teh, "A fast learning algorithm for deep belief nets," Neural computation, vol. 18, pp. 1527-1554, 2006.

[11] X. Ma, Z. Tao, Y. Wang, H. Yu, and Y. Wang, "Long short-term memory neural network for traffic speed prediction using remote microwave sensor data," Transportation Research Part C: Emerging Technologies, vol. 54, pp. 187-197, 2015.

[12] R. Fu, Z. Zhang and L. Li, "Using LSTM and GRU neural network methods for traffic flow prediction," in Chinese Association of Automation (YAC), Youth Academic Annual Conference, 2016, pp. 324-328.

[13] J. Guo, Z. Xie, Y. Qin, L. Jia, and Y. Wang, "Short-Term Abnormal Passenger Flow Prediction Based on the Fusion of SVR and LSTM," IEEE Access, vol. 7, pp. 42946-42955, 2019.

[14] Q. Tang, M. Yang and Y. Yang, "ST-LSTM: A Deep Learning Approach Combined Spatio-Temporal Features for Short-Term Forecast in Rail Transit," Journal of Advanced Transportation, vol. 2019, pp. 1-8, 2019.

[15] H. Li, Y. Wang, X. Xu, L. Qin, and H. Zhang, "Short-term passenger flow prediction under passenger flow control using a dynamic radial basis function network," Applied Soft Computing, vol. 83, pp. 105620, 2019.

[16] X. Ma, Z. Dai, Z. He, J. Ma, Y. Wang, and Y. Wang, "Learning traffic as images: a deep convolutional neural network for large-scale transportation network speed prediction," Sensors, vol. 17, pp. 818, 2017.

[17] S. Kim, J. Kang, M. Lee, and S. Song, "DeepTC: ConvLSTM Network for Trajectory Prediction of Tropical Cyclone using Spatiotemporal Atmospheric Simulation Data," in NIPS 2018 Workshop Spatiotemporal, 2018, pp. 1-5.

[18] L. Y., D. Y., K. W., L. Z., and W. F., "Traffic Flow Prediction with Big Data: A Deep Learning Approach," IEEE Transactions on Intelligent Transportation Systems, vol. 16, pp. 865-873, 2015.

[19] Y. Liu, Z. Liu and R. Jia, "DeepPF: A deep learning based architecture for metro passenger flow prediction," Transportation Research Part C: Emerging Technologies, vol. 101, pp. 18-34, 2019.

[20] K. He, X. Zhang, S. Ren, and J. Sun, "Deep residual learning for image recognition," in Proceedings of the IEEE conference on computer vision and pattern recognition, 2015, pp. 770-778.

[21] P. Liu, Y. Zhang, D. Kong, and B. Yin, "Improved Spatio-Temporal Residual Networks for Bus Traffic Flow Prediction," Applied Sciences, vol. 9, pp. 615, 2019.

[22] J. Zhang, Y. Zheng and D. Qi, "Deep spatio-temporal residual networks for citywide crowd flows prediction," in Thirty-First AAAI Conference on Artificial Intelligence, 2017, pp. 1-7.

[23] L. Li, L. Qin, X. Qu, J. Zhang, Y. Wang, and B. Ran, "Day-ahead traffic flow forecasting based on a deep belief network optimized by the multi-objective particle swarm algorithm," Knowledge-Based Systems, vol. 172, pp. 1-14, 2019.

[24] L. Li, J. Zhang, Y. Wang, and B. Ran, "Missing value imputation for traffic-related time series data based on a multi-view learning method," IEEE Transactions on Intelligent Transportation Systems, vol. 20, pp. 2933-2943, 2018.

[25] B. Yu, Y. Lee and K. Sohn, "Forecasting road traffic speeds by considering area-wide spatio-temporal dependencies based on a graph convolutional neural network (GCN)," Transportation Research Part C: Emerging Technologies, vol. 114, pp. 189-204, 2020.

[26] T. Bogaerts, A. D. Masegosa, J. S. Angarita-Zapata, E. Onieva, and P. Hellinckx, "A graph CNN-LSTM neural network for short and long-term traffic forecasting based on trajectory data," Transportation Research Part C: Emerging Technologies, vol. 112, pp. 62-77, 2020.

[27] L. Zhao, Y. Song, C. Zhang, Y. Liu, P. Wang, T. Lin, M. Deng, and H. Li, "T-GCN: A Temporal Graph Convolutional Network for Traffic Prediction," IEEE Transactions on Intelligent Transportation Systems, 2019 (Online).

[28] Z. Cui, K. Henrickson, R. Ke, and Y. Wang, "Traffic graph convolutional recurrent neural network: A deep learning framework for network-scale traffic learning and forecasting," IEEE Transactions on Intelligent Transportation Systems, 2019 (Online).

[29] J. Li, H. Peng, L. Liu, G. Xiong, B. Du, H. Ma, L. Wang, and M. Z. A. Bhuiyan, "Graph CNNs for Urban Traffic Passenger Flows Prediction," in 2018 IEEE SmartWorld, Ubiquitous Intelligence & Computing, Advanced & Trusted Computing, Scalable Computing & Communications, Cloud & Big Data Computing, Internet of People and Smart City Innovation, 2018, pp. 29-36.

[30] Han, J. Wang, Ren, Gao, and Chen, "Predicting Station-Level Short-Term Passenger Flow in a Citywide Metro Network Using Spatiotemporal Graph Convolutional Neural Networks," International Journal of Geo-Information, vol. 8, pp. 243, 2019.

[31] C. Szegedy, W. Liu, Y. Jia, P. Sermanet, S. Reed, D. Anguelov, D. Erhan, V. Vanhoucke, and A. Rabinovich, "Going deeper with convolutions," in Proceedings of the IEEE conference on computer vision and pattern recognition, 2015, pp. 1-9.

[32] X. Glorot and Y. Bengio, "Understanding the difficulty of training deep feedforward neural networks," in Proceedings of the thirteenth international conference on artificial intelligence and statistics, 2010, pp. 249-256.

[33] K. He, X. Zhang, S. Ren, and J. Sun, "Identity mappings in deep residual networks," in European conference on computer vision, 2016, pp. 630-645.

[34] T. N. Kipf and M. Welling, "Semi-supervised classification with graph convolutional networks," arXiv preprint arXiv:1609.02907, 2016.

[35] Z. Wu, S. Pan, F. Chen, G. Long, C. Zhang, and P. S. Yu, "A comprehensive survey on graph neural networks," arXiv preprint arXiv:1901.00596, 2019.

[36] G. Li, M. Müller, A. Thabet, and B. Ghanem, "Can GCNs Go as Deep as CNNs?" arXiv preprint arXiv:1904.03751, 2019.

[37] Q. Li, Z. Han and X. Wu, "Deeper insights into graph convolutional networks for semi-supervised learning," in Thirty-Second AAAI Conference on Artificial Intelligence, 2018, pp.1-8.

[38] D. Bahdanau, K. Cho and Y. Bengio, "Neural machine translation by jointly learning to align and translate," arXiv preprint arXiv:1409.0473, 2014.

[39] Z. Zhang, M. Li, X. Lin, Y. Wang, and F. He, "Multistep speed prediction on traffic networks: A deep learning approach considering spatio-temporal dependencies," Transportation Research Part C: Emerging Technologies, vol. 105, pp. 297-322, 2019.

[40] Q. Liu, B. Wang and Y. Zhu, "Short-Term Traffic Speed Forecasting Based on Attention Convolutional Neural Network for Arterials," Computer-Aided Civil and Infrastructure Engineering, vol. 33, pp. 999-1016, 2018.

[41] Y. Wu, H. Tan, L. Qin, B. Ran, and Z. Jiang, "A hybrid deep learning based traffic flow prediction method and its understanding," Transportation Research Part C: Emerging Technologies, vol. 90, pp. 166-180, 2018.

[42] M. Abadi, P. Barham, J. Chen, Z. Chen, A. Davis, J. Dean, M. Devin, S. Ghemawat, G. Irving, and M. Isard, "Tensorflow: A system for large-scale machine learning," in 12th Symposium on Operating Systems Design and Implementation, 2016, pp. 265-283.

[43] M. J. Norusis and M. J. Norusis, SPSS for Windows: base system user's guide, release 6.0. MI, USA: SPSS Incorporated, 1993, pp. 1-828.

[44] F. Pedregosa, G. Varoquaux, A. Gramfort, V. Michel, B. Thirion, O. Grisel, M. Blondel, P. Prettenhofer, R. Weiss, and V. Dubourg, "Scikit-learn: Machine learning in Python," Journal of machine learning research, vol. 12, pp. 2825-2830, 2011.

[45] S. Xingjian, Z. Chen, H. Wang, D. Yeung, W. Wong, and W. Woo, "Convolutional LSTM network: A machine learning approach for precipitation nowcasting," in Advances in neural information processing systems, 2015, pp. 802-810.

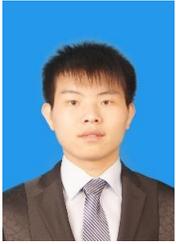
**Jinlei Zhang** was born in Hebei Province, China. He is a PhD candidate in Beijing Jiaotong University and a visiting PhD student in University of Washington.

His research interests include machine learning, deep learning, traffic data mining and applications, dynamic traffic modeling and management.

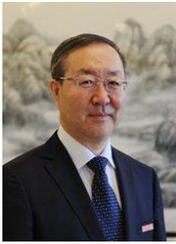
**Feng Chen** received his B.S. degree, M.A. degree, Ph.D. degree in railway engineering from Beijing Jiaotong University, Beijing, China, in 1983, 1990 and 2007 respectively. He is now working as a professor in Beijing Jiaotong University and the president of Chang'an University.

His research interests include passenger flow management, traffic data mining and application for urban rail transit.

Professor Chen won the first prize of national science and technology progress as the second accomplisher in 2017.

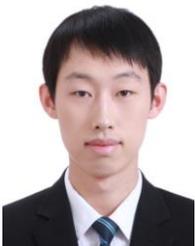
**Zhiyong Cui** received a B.S. degree in Software Engineering from Beihang University (2012) and a M.S. degree in Software Engineering and Microelectronics from Peking University (2015). He is currently working toward the Ph.D. degree in Civil and Environmental Engineering at University of Washington.

His research interests include deep learning, machine learning, urban computing, connected vehicles, autonomous driving, and transportation data science. He serves as a member of Transportation Research Board (TRB) Standing Committee on Intelligent Transportation Systems (AHB15) and Standing Committee on Geospatial Data Acquisition Committee (AFB80).

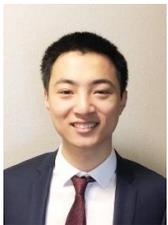
**Yinan Guo** is an undergraduate student pursuing a double degree in Mathematics and Informatics at the University of Washington. His current research interest lies in the interplay among mathematics, information science and computer science.

His interests in mathematics, science and technology rooted in his high school study in New Hampshire. Since then, he has received one international award and three patents for his invention.

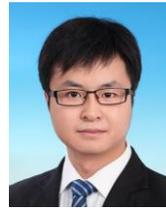
**Yadi Zhu** was born in Henan Province, China. He received the Ph.D. degree from the School of Civil Engineering, Beijing Jiaotong University in 2019. He currently holds a postdoctoral position with the School of Engineering, Center for Advanced Infrastructure and Transportation, Rutgers University.

His research interests are transportation planning, including pedestrian flow modeling, pedestrian simulation research, and transportation demand analysis and forecasting.